\documentclass[sigconf]{acmart}

\usepackage{booktabs} 
\usepackage{multirow}
\usepackage{enumitem}

\setcopyright{rightsretained}

\acmConference[ACM RecSys Poster]{}{August 2017}{}
\acmYear{2017}
\copyrightyear{2017}

\acmPrice{00.00}

\begin{document}
\title{pyRecLab: A Software Library for Quick Prototyping of Recommender Systems}

\author{Gabriel Sepulveda}
\affiliation{%
  \institution{Pontificia Universidad Catolica de Chile}
  \city{Santiago}
  \state{Chile}
}
\email{grsepulveda@uc.cl}

\author{Vicente Dominguez}
\affiliation{%
    \institution{Pontificia Universidad Catolica de Chile}
    \city{Santiago}
    \state{Chile}
}
\email{vidominguez@uc.cl}

\author{Denis Parra}
\affiliation{%
    \institution{Pontificia Universidad Catolica de Chile}
    \city{Santiago}
    \state{Chile}
}
\email{dparra@ing.puc.cl}

\renewcommand{\shortauthors}{Sepulveda et al.}

\begin{abstract}
This paper introduces pyRecLab, a software library written in C++ with Python bindings which allows to quickly train, test and develop recommender systems. Although there are several software libraries for this purpose, only a few let developers to get quickly started with the most traditional methods, permitting them to try different parameters and approach several tasks without a significant loss of performance. Among the few libraries that have all these features, they are available in languages such as Java, Scala or C\#, what is a disadvantage for less experienced programmers more used to the popular Python programming language. In this article we introduce details of pyRecLab, showing as well performance analysis in terms of error metrics (MAE and RMSE) and train/test time. We benchmark it against the popular Java-based library LibRec, showing similar results. We expect programmers with little experience and people interested in quickly prototyping recommender systems to be benefited from pyRecLab.
\end{abstract}

%
%


\keywords{Recommender Systems, Software Development, Recommender Library, Python Library}

\maketitle

\section{Introduction}

When software developers face the challenge of learning about recommender systems (RecSys), developing a RecSys for the first time, or quickly prototyping a recommender to test available data, a reasonable option to get started is using an existent software library. Nowadays, it is possible to find several libraries in different programming languages, being among of the most popular ones MyMedialite~\cite{mymedialite}, LensKit~\cite{lenskit}, LibRec~\cite{librec}, lightfm~\cite{lightfm} and rrecsys~\cite{rrecsys}.



%
\begin{figure}[t!]
\includegraphics[width=\linewidth]{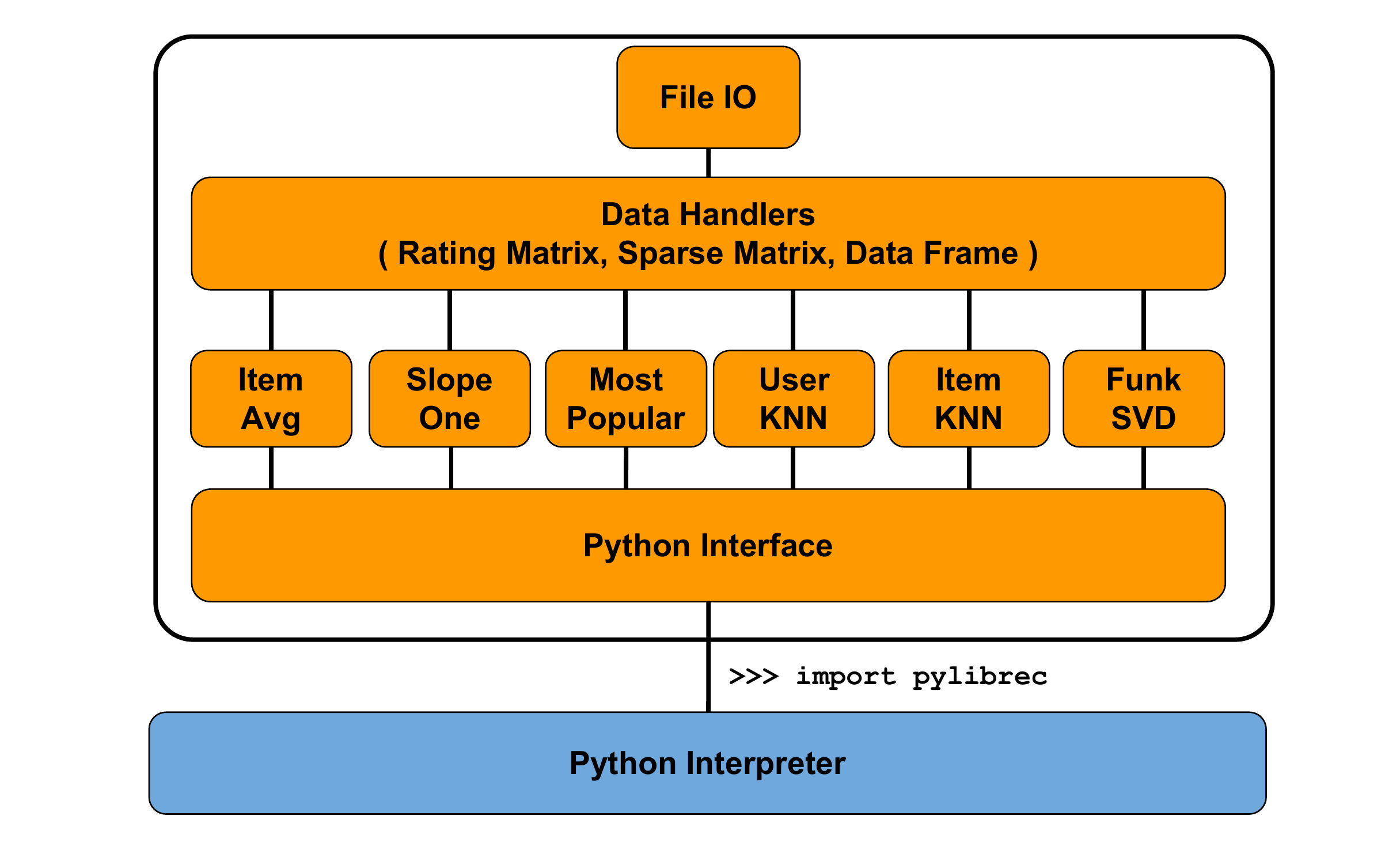}
\caption{\texttt{pyRecLab} architecture.}
\label{fig:pyreclab-arch}
\end{figure}

While the aforementioned tools have documentation, implement several methods, and present most of the common functionality required to develop and evaluate a recommendation system, all of them miss some type of functionality or algorithm which hinder specially newcomers. In particular, while teaching for three years a graduate course on Recommender Systems during the Fall Semester (2014-2016) at a Department of Computer Science, most students have found recurrent difficulties in using existent tools to finish an introductory assignment. The assignment is related to tasks such as rating prediction and item recommendation to specific users, using well-known collaborative filtering methods such as User K-NN, Item K-NN, Slope One and FunkSVD \cite{parra2013recommender}. Some of the problems found were: (a) the lack of implementation of certain methods in some libraries, (b) poor train/test time performance under medium-sized datasets (such as Rrecsys which does not implement sparse matrices), (c) lack of functionality which is typical in a recommendation setting, such us suggesting a list of items given a specific user ID, (d) difficulties to change parameters in certain models, and (e) students' lack of familiarity with certain programming languages such as Java or C\#. While Java is the most popular language based on several rankings, it is also the case that Python is the most popular introductory teaching language in the U.S. since 2004 \cite{guo2014python} as well as the one with largest growth in the latest 5 years based on the PYPL ranking\footnote{http://pypl.github.io/PYPL.html}.

For these reasons, we developed \emph{pyRecLab}\footnote{Documentation and code samples at https://github.com/gasevi/pyreclab}. We wrote it in C++ with Python bindings, in order to facilitate its adoption among new programmers familiar with Python, but also offering an appropriate performance when dealing with larger datasets. We implemented most of the foundational recommendation methods for rating prediction and recommendation. Moreover, users can easily change parameters to understand their effect and they can also produce recommendations given a specific user ID.

\section{Other Recommendation Libraries}

\textbf{MyMediaLite}\cite{mymedialite}: It implements several recommendation algorithms, supporting explicit and implicit feedback, as well as context-aware methods. It also allows evaluation with metrics such as MAE, RMSE, prec@N, and nDCG \cite{parra2013recommender}. Many of it functionalities are available from command line; however, to integrate it with other software it is necessary to program in languages like C\# or F\#, which is difficult for many newcomer Python developers.

\textbf{Lenskit}\cite{lenskit}: A popular library which provides all basic collaborative filtering methods for predicting ratings (User/Item KNN, Slope One and FunkSVD). It is developed in Java, which could be an entry barrier for new programmers who are mostly familiar with Python.

\textbf{LibRec}\cite{librec}: Just like MyMediaLite and Lenskit, a well developed library in terms of algorithms implemented and the metrics available for evaluation. However, documentation is not as good as Lenskit and since it is implemented in Java, it also raises the barrier for new programmers.

\textbf{Lightfm}\cite{lightfm}: This library implements several matrix factorization algorithms for both implicit and explicit feedback. It also has an interface for Python, facilitating its use to several developers. However, it does not implement basic traditional recommender algorithms (User/Item KNN, slope One), so it is not advisable for introductory teaching purposes.

\textbf{Rrecsys}\cite{rrecsys}: This tool gets the closest to pyRecLab in terms of easy-of-use, quick prototyping and educational purposes. It is written in R language. However, it has two main weaknesses: it misses some traditional algorithms (like Slope One) and it is limited in terms of the amount of data it can process, since it does not support sparse matrices.

\begin{table}[t!]
\centering
\caption{\texttt{pyRecLab} vs. \texttt{LibRec} on MovieLens 100K data.}
\label{tab:pyreclab-mae-rmse}
\scalebox{1}{
\renewcommand{\arraystretch}{0.8}%
\begin{tabular}{@{}lllll@{}}
\toprule
\multirow{2}{*}{} & \multicolumn{2}{c}{MAE}                                   & \multicolumn{2}{c}{RMSE}                                  \\ \cmidrule(l){2-5}
                           & \multicolumn{1}{c}{pyRecLab} & \multicolumn{1}{c}{LibRec} & \multicolumn{1}{c}{pyRecLab} & \multicolumn{1}{c}{LibRec} \\ \midrule
UserAvg                    & 0.850191                     & 0.850191                   & 1.062995                     & 1.062995                   \\ \midrule
ItemAvg                    & 0.827568                     & 0.827568                   & 1.033411                     & 1.033411                   \\ \midrule
SlopeOne                   & 0.748552                     & 0.748299                   & 0.952795                     & 0.952460                   \\ \midrule
User KNN                   & 0.754816                     & 0.755361                   & 0.962355                     & 0.966395                   \\ \midrule
Item KNN                   & 0.749316                     & 0.748354                   & 0.953637                     & 0.953433                   \\ \midrule
Funk SVD                   & 0.732820                     & 0.731986                   & 0.925390                     & 0.923978                   \\ \bottomrule
\end{tabular}
}
\end{table}

\section{Design and Implementation}
Figure \ref{fig:pyreclab-arch}, shows the main modules of pyRecLab. At the bottom, the blue block represents the Python interpreter, which  loads the methods and data structures when importing the PyRecLab module.
At the top, in orange, all the sub-modules of the library:

\begin{itemize}[leftmargin=*]
\item \textbf{File IO}. This component allows data input/output by means of reading from text files, as well as writing output recommendations in txt and json formats. It allows great flexibility in terms of input file formats (csv, tsv) as well as allowing the user to specify what to file columns represent.

\item \textbf{Data handlers}. This module implements several data structures, which allow a homogeneous access to the ratings. It grants a good level of independence from the original format from which data were read, with a high level of abstraction. These data structures will be directly used by the recommendation algorithms for the processing, storage and generation of output data.

\item \textbf{Recommendation Algorithms}. Under the Data handlers block, there are a number of contiguous blocks representing the recommendation algorithms. Algorithms for rating prediction and recommendation are: Item Average, Slope One, User KNN, Item KNN and Funk SVD. On the other hand, Most Popular is only used to generate recommendations.

\item \textbf{Python Interface}. This module represents the interface between the recommendation algorithms and the Python interpreter. It was developed in C++, and since we aimed at maintaining an appropriate level of code readability, we decided to use the Python/C API rather than Cython for implementation. This allows us to define low-level structures in C++ language with a direct mapping with objects handled by the Python interpreter. In this way, we have defined a data type for each of the recommendation algorithms, which can be instantiated directly from the Python interpreter.

\end{itemize}

\section{Results \& Conclusion}

\begin{figure}[t!]
\includegraphics[width=0.9\linewidth]{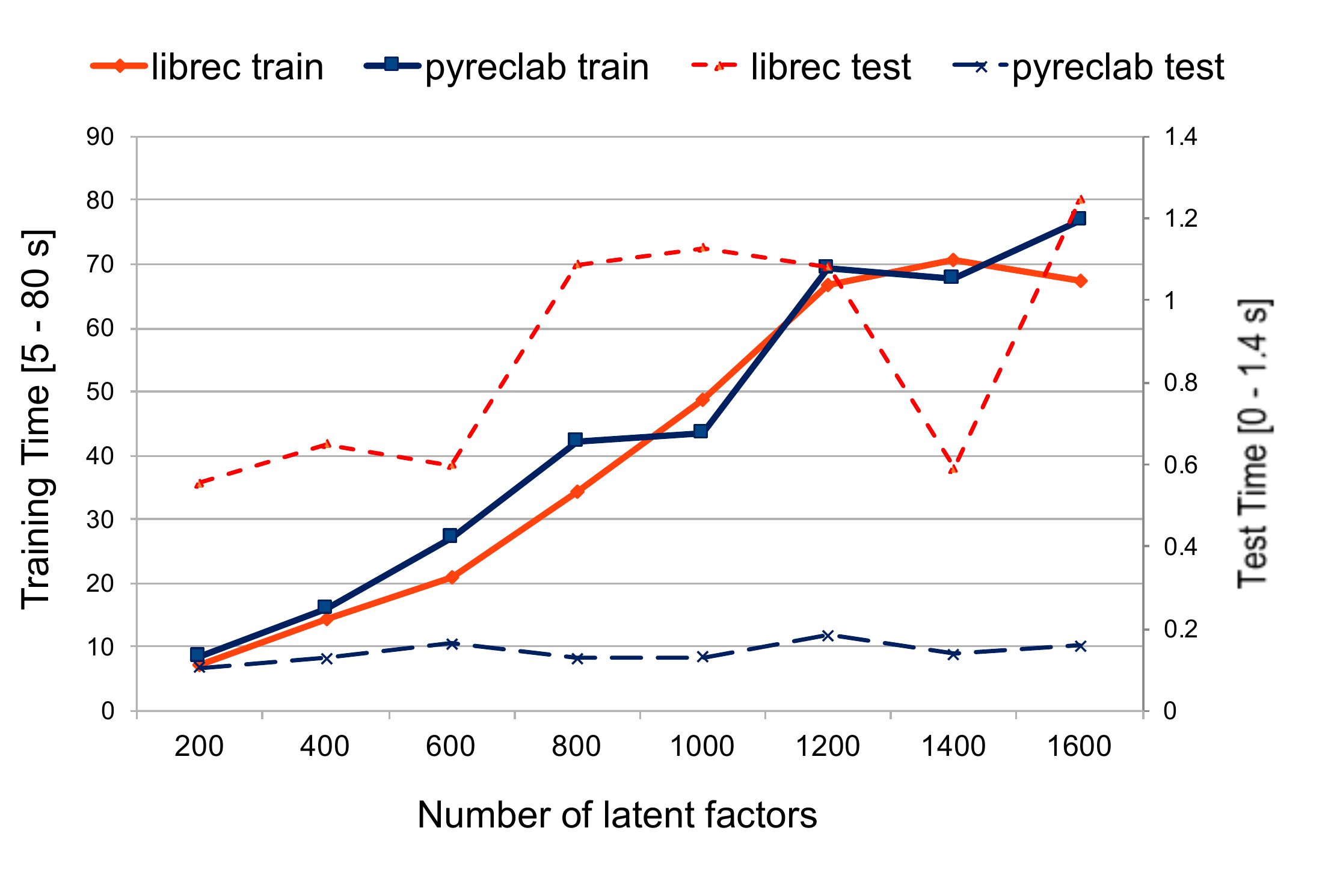}
\caption{\texttt{pyRecLab} vs. \texttt{LibRec} on time performance.}
\label{fig:pyreclab-time}
\end{figure}

To check the performance of {\em pyRecLab}, we tested it against the popular library LibRec \cite{librec} in terms of error and train/test time.

{\bf Prediction Results}. MAE and RMSE results of rating prediction over Movielens 100K dataset are shown in Table \ref{tab:pyreclab-mae-rmse}. Differences are very small to {\em LibRec}, showing that {\em pyRecLab} can reproduce results of a mature recommender library.

{\bf Time Performance}. Although the results vary depending on the method, Figure \ref{fig:pyreclab-time} shows train/test performance using FunkSVD. While both libraries perform similarly in training phase, pyRecLab performs faster in testing time at different number of latent factors.

Summarizing, we have introduced PyRecLab, a library for recommender systems which combines the performance of C++ in its implementation with the versatility of Python for easy-of-use. We expect to add algorithms and recommendations metrics, as well as new code samples to facilitate its widespread adoption.

\bibliographystyle{ACM-Reference-Format}
\bibliography{sigproc}

\end{document}